\documentclass{pasj01}
\twocolumn
\usepackage{multirow}
\usepackage{natbib}
\Received{$\langle$2018 February 9$\rangle$}
\Accepted{$\langle$2018 March 28$\rangle$}
\Published{$\langle$publication date$\rangle$}

\begin{document}
\title{Supermassive black holes with higher Eddington ratios preferentially form in gas-rich galaxies}
\author{Takuma \textsc{Izumi},\altaffilmark{1,}$^{\dag,*}$ 
}
\altaffiltext{1}{National Astronomical Observatory of Japan, 2-21-1 Osawa, Mitaka, Tokyo 181-8588, Japan }
\altaffiltext{$\dag$}{NAOJ Fellow}
\email{takuma.izumi@nao.ac.jp}

\maketitle

\begin{abstract}
The Eddington ratio ($\lambda_{\rm Edd}$) of supermassive black holes (SMBHs) 
is a fundamental parameter that governs the cosmic growth of SMBHs. 
Although gas mass accretion onto SMBHs is sustained 
when they are surrounded by large amounts of gas, 
little is known about the molecular content of galaxies, particularly those hosting super-Eddington SMBHs 
($\lambda_{\rm Edd} > 1$: the key phase of SMBH growth). 
Here, we compiled reported optical and $^{12}$CO(1--0) data of local quasars to characterize their hosts. 
We found that higher $\lambda_{\rm Edd}$ SMBHs tend to reside in gas rich 
(i.e., high gas mass to stellar mass fraction = $f_{\rm gas}$) galaxies. 
We used two methods to make this conclusion: one uses black hole mass 
as a surrogate for stellar mass by assuming a local co-evolutionary relationship, 
and the other directly uses stellar masses estimated from near-infrared observations. 
The $f_{\rm gas}$--$\lambda_{\rm Edd}$ correlation we found concurs 
with the cosmic decreasing trend in $\lambda_{\rm Edd}$, 
as cold molecular gas is primarily consumed by star formation. 
This correlation qualitatively matches predictions of recent semi-analytic models 
about the cosmic downsizing of SMBHs as well. 
As the gas mass surface density would eventually be a key parameter controlling mass accretion, 
we need high-resolution observations to identify further 
differences in the molecular properties around super-Eddington and sub-Eddington SMBHs. 
\end{abstract}

\section{Introduction}\label{sec1}
The huge amount of energy emitted by active galactic nuclei (AGN) 
and high redshift quasars is commonly ascribed to gas accretion by supermassive black holes (SMBHs). 
The co-evolution of SMBHs and galaxies is supported by the tight correlations 
between SMBH mass ($M_{\rm BH}$) and some properties of their host galaxies 
such as bulge stellar mass and stellar velocity dispersion 
\citep[][for a review]{2013ARA&A..51..511K}, 
as well as the similarity of the global history 
of star formation and AGN activity \citep[][for a review]{2014ARA&A..52..415M}. 

The evolutionary stage of an AGN/SMBH can be characterized by its Eddington ratio ($\lambda_{\rm Edd}$)
\footnote{The ratio of AGN bolometric luminosity ($L_{\rm Bol}$) to its Eddington luminosity ($L_{\rm Edd}$), 
where $L_{\rm Edd}$ [erg s$^{-1}$] = 1.26 $\times$ 10$^{38}$ $M_{\rm BH}$ [$M_\odot$].}. 
\citet{2004A&A...420L..23K} indicated that the cosmic SMBH growth 
by gas accretion is dominated by the super-Eddington phase (i.e., $\lambda_{\rm Edd} > 1$), 
rather than the sub-Eddington phase common among low redshift AGN \citep[e.g.,][]{2012ApJ...746..169S}. 
Such a mass accretion process would be sustained by large amounts of cold molecular gas 
in the central regions of galaxies \citep[e.g.,][]{2008ApJ...681...73K,2008A&A...491..441V}. 
Indeed, recent millimeter/submillimeter observations have shown 
that the {\it mass surface density} of a sub-kpc scale 
dense molecular gas structure is a key parameter controlling 
the mass accretion rate of an SMBH \citep[e.g.,][]{2011ApJS..195...23M,2016ApJ...827...81I,2017MNRAS.470.1570H}, 
although those works focused mostly on sub-Eddington objects. 
\citet{2005ApJ...625...78H} showed that infrared-luminous quasars tend 
to possess super-Eddington SMBHs. 
Therefore it is plausible that super-Eddington objects occur 
in young starburst galaxies \citep{2012ApJ...750...92X}, 
corresponding to the early/gas-rich phase of galaxy evolution, 
which is indeed expected in a number of theoretical/numerical models \citep[e.g.,][]{2006ApJS..163....1H}. 

The space density of AGN over cosmic time 
can also be a stringent constraint on SMBH/galaxy evolutionary models. 
In this sense, AGN {\it downsizing} or {\it anti-hierarchical evolution} is another key trend that we should consider: 
both ultraviolet \citep[e.g.,][]{2009MNRAS.399.1755C,2012ApJ...756..160I} 
and X-ray \citep[e.g.,][]{2005A&A...441..417H,2014ApJ...786..104U} 
observations have revealed that the space densities of luminous AGN peak at a redshift higher than those of faint AGN. 
If the AGN luminosity reflects $M_{\rm BH}$, 
the downsizing indicates that the more massive systems formed at earlier cosmic times. 
This seems to contradict the hierarchical structure formation in the cold dark matter (CDM) universe. 
However, if a super-Eddington SMBH, which would occur in the dominant 
phase of mass accumulation \citep{2004A&A...420L..23K}, 
is characteristic of a specific environment that is common at high-$z$, this trend could be natural. 

In this Letter, we first compile galaxy-scale $^{12}$CO(1--0) emission line measurements 
of low redshift ($z \lesssim 0.3$) quasars, for which rich multi-wavelength data are available, 
to estimate their global cold molecular mass ($M_{\rm gas}$). 
The relevance of the gas mass fraction to $\lambda_{\rm Edd}$ is then explored in two different ways. 
Hereafter, we define the gas mass fraction as 
$f_{\rm gas} \equiv M_{\rm gas}/M_\star$, where $M_\star$ denotes the galaxy total stellar mass. 
The standard $\Lambda$-CDM cosmology with $H_0$ = 70 km s$^{-1}$ Mpc$^{-1}$, 
$\Omega_{\rm M}$ = 0.3, $\Omega_{\rm \Lambda}$ = 0.7 is adopted throughout this work.

\section{Method and Data}\label{sec2} 
\begin{figure*}
\begin{center}
\includegraphics[scale=0.475]{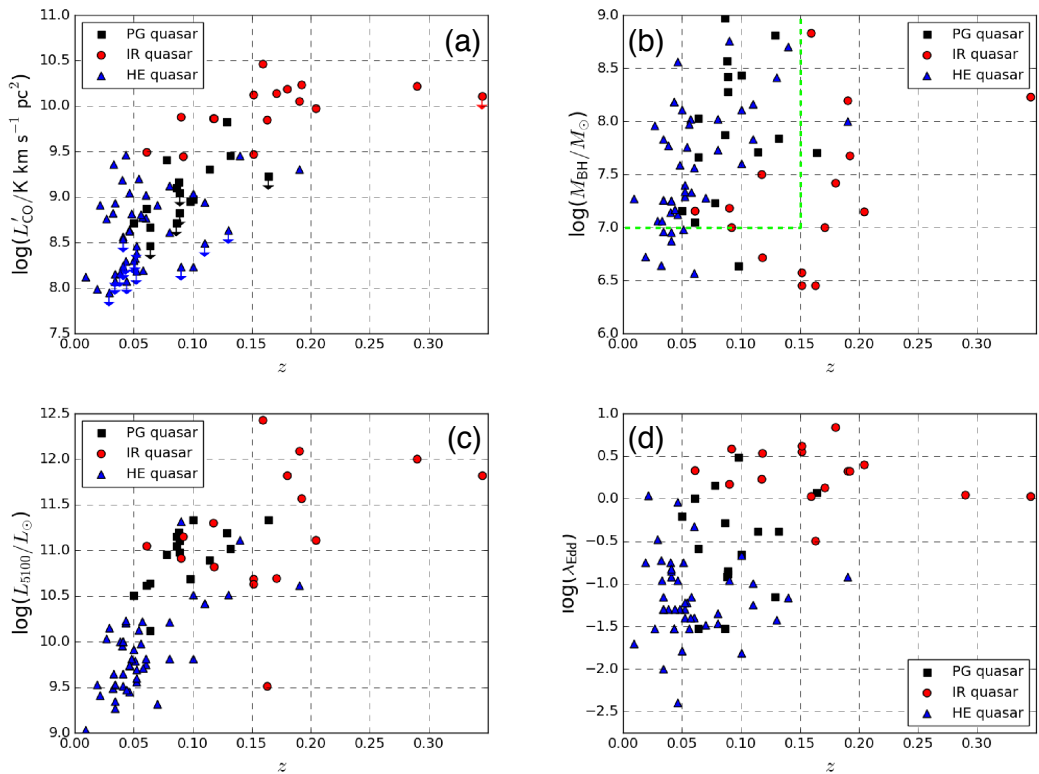}
\end{center}
\caption{
The distributions of (a) $^{12}$CO(1--0) line luminosity, 
(b) $M_{\rm BH}$, (c) $L_{\rm 5100}$, and (d) $\lambda_{\rm Edd}$, 
of our samples as functions of the redshift. 
PG quasars (black squares), IR quasars (red circles), 
and HE quasars (blue triangles) are shown. 
The dashed green lines in (b) defines our {\it subsample}, 
i.e., $M_{\rm BH} \gtrsim 10^7~M_\odot$ $\cap$ $z < 0.15$ 
(see details in Section 3). 
}
\label{fig1}
\end{figure*}

Our immediate objective is to examine the connection 
between $\lambda_{\rm Edd}$ and $f_{\rm gas}$ 
to characterize the nature of the host galaxies of super-Eddington SMBHs. 
For this purpose, we applied two methods. 
The first {\it indirect} method uses $M_{\rm BH}$ as a surrogate for $M_\star$ 
by assuming that the local co-evolutionary relationship also holds in our samples. 
The second {\it direct} method faithfully uses $M_\star$ derived 
from near-infrared (NIR; $H$- or $K$-band) imaging studies to estimate $f_{\rm gas}$. 
Although the latter is more straightforward, 
the number of sample quasars decreases (Section 2.4), 
as high-resolution NIR observations are required to separate AGN light from stellar light. 
Considering the above methods, we compiled 
multi-wavelength literature data for low redshift quasars as described below. 
The relevant properties of our samples are visualized in Figure \ref{fig1}. 

\subsection{PG quasars} 
As our base sample, we first compiled optically luminous local quasars 
(5100 {\AA} continuum luminosity, $L_{\rm 5100} \sim 10^{44-45}$ erg s$^{-1}$) 
from the Palomar-Green (PG) Bright Quasar Survey 
\citep{1983ApJ...269..352S} with galaxy-scale $^{12}$CO(1--0) measurements 
\citep{2001AJ....121.1893E,2006AJ....132.2398E,2009AJ....138..262E,2003ApJ...585L.105S}. 
All of the $^{12}$CO(1--0) observations were made 
with the Owens Valley Radio Observatory (OVRO) interferometer, 
providing angular resolutions of $\sim 4\arcsec-5\arcsec$ (corresponding to $\sim 4-5$ kpc). 
These still cover a large fraction of their galaxy-scale $^{12}$CO(1--0) emissions, 
given that low-$J$ CO emissions typically come from the innermost $\lesssim 5$ kpc regions 
of both nearby Seyfert-class AGN and high redshift quasars 
\citep[e.g.,][]{2011ApJ...739L..34W,2015ApJ...811...39I}. 

We applied the Galactic CO-to-$M_{\rm gas}$ conversion factor of 
$\alpha_{\rm CO}$ = 4 $M_\odot$ (K km s$^{-1}$ pc$^2$)$^{-1}$ \citep{2013ARA&A..51..207B} 
to the line luminosities $L'_{\rm CO}$, 
as the $^{12}$CO(1--0) data cover both the circumnuclear $\lesssim 100$ pc region 
where lower $\alpha_{\rm CO}$ may apply, 
and rather quiescent $>$kpc scale regions where 
interstellar medium with properties similar to those of our Galaxy would be expected. 
Although the actual value of $\alpha_{\rm CO}$ would vary with the source, 
we use the above $\alpha_{\rm CO}$ throughout this work to maintain consistency among the samples. 
The typical uncertainty of $\alpha_{\rm CO}$ is $\sim 0.3$ dex \citep{2013ARA&A..51..207B}. 

$M_{\rm BH}$ of those PG quasars was estimated with the so-called 
single-epoch method \citep[e.g.,][]{2000ApJ...533..631K,2006ApJ...644..133B} 
that uses H$\beta$ FWHM and $L_{\rm 5100}$, 
with a virial factor of 1, which are collected from 
\citet{2005ApJ...625...78H} and \citet{2015ApJ...806...22D}. 
A typical uncertainty of $M_{\rm BH}$ based on this method is $\sim 0.3$ dex \citep{2008ApJ...680..169S}, 
which stems from the dispersion in the reverberation-mapped $M_{\rm BH}$ calibration. 
The AGN bolometric luminosities ($L_{\rm Bol}$) are also derived from $L_{\rm 5100}$, 
by applying a bolometric correction factor of 9 \citep{2000ApJ...533..631K}, 
which has the typical uncertainty of $\sim 0.3$ dex. 
Therefore, we expect that the uncertainty of $\lambda_{\rm Edd}$ is $\sim 0.4$ dex. 
These procedures were applied to the following samples as well. 
After reclassifying three PG quasars as IR quasars (Section 2.2) due to their high IR luminosity, 
16 quasars were categorized in this PG quasar sample.

\subsection{IR quasars} 
Local quasars showing high IR luminosity ($L_{\rm 8-1000\mu m} > 10^{12}~L_\odot$; IR quasar) 
compiled in \citet{2005ApJ...625...78H} were matched with 
the $^{12}$CO(1--0) measurements with the IRAM 30 m telescope presented in \citet{2012ApJ...750...92X}. 
The achieved angular resolutions were $\sim 22\arcsec$, 
so their galaxy-scale $^{12}$CO(1--0) emissions should be fully recovered. 
After adding the three reclassified PG quasars, our IR quasar sample consisted of 16 quasars 
to widen the range of $\lambda_{\rm Edd}$, 
as they tend to show $\lambda_{\rm Edd} \gtrsim 1$ \citep{2005ApJ...625...78H}. 
$M_{\rm BH}$, $M_{\rm gas}$, $\lambda_{\rm Edd}$ of the sample IR quasars 
are calculated in the same manner as for the PG quasar sample. 
Note that, although one would expect that the IR quasars possess huge amounts of $M_{\rm gas}$, 
which is actually the case as judged from Figure \ref{fig1}, 
their $f_{\rm gas}$ is not so obvious. 

\subsection{HE quasars} 
Our sample was further extended toward the optically 
lower luminosity regime ($L_{\rm 5100} \lesssim 10^{44}$ erg s$^{-1}$) 
by adding quasars from the Hamburg/ESO (HE) Survey \citep{2000A&A...358...77W}. 
Systematic measurements of their $^{12}$CO(1--0) emission line with the IRAM 30 m telescope 
were performed by \citet{2007A&A...470..571B} and \citet{2017MNRAS.470.1570H}. 
Then, we matched their data with the $M_{\rm BH}$ and $L_{\rm 5100}$ (or $\lambda_{\rm Edd}$) 
data presented in \citet{2010A&A...516A..87S} 
and \citet{2014MNRAS.443..755H,2017MNRAS.470.1570H} to construct the HE quasar sample. 
There are 44 quasars in this sample. 

\subsection{Stellar mass}
\begin{figure}
\begin{center}
\includegraphics[scale=0.3]{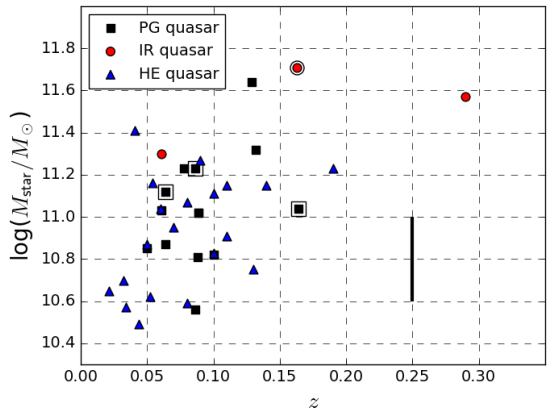}
\end{center}
\caption{
The $M_\star$ distribution of our sample of PG (black squares), IR (red circles), 
and HE (blue triangles) quasars as a function of the redshift. 
These were measured with NIR ($H$- or $K$-band) photometric observations. 
The black bar indicates the expected uncertainty due to the assumed mass-to-light ratio 
\citep[$\sim 0.2$ dex;][]{2016ApJ...819L..27Z}. 
The double symbols denote the secondary data, for which $M_*$ was measured with $V$-band photometry. 
}
\label{fig2}
\end{figure}

$M_\star$ data for 11 of our PG quasar samples \citep[69\%,][]{2016ApJ...819L..27Z}, 
two of the IR quasar samples \citep[13\%,][]{2016ApJ...819L..27Z}, 
and 20 of the HE quasar samples \citep[45\%,][]{2004MNRAS.352..399J,2014A&A...561A.140B}, are available from the literature. 
The $M_*$-estimates were based on high-resolution $H$- or $K$-band photometry 
with the {\it Hubble Space Telescope} or ground-based AO-assisted systems. 
Mass-to-luminosity ratios ($M/L$) of -0.017 ($H$) and -0.08 ($K$) 
on a logarithmic scale were assumed to maintain consistency with \citet{2016ApJ...819L..27Z}. 
A typical uncertainty in NIR-based $M_\star$ due to the assumed 
$M/L$ is $\sim 0.2$ dex \citep{2016ApJ...819L..27Z}. 
Their $M_\star$ range as $\sim 10^{10.6}-10^{11.6}~M_\odot$ (PG quasar), 
$\sim 10^{11.3}-10^{11.6}~M_\odot$ (IR quasar), 
and $\sim 10^{10.5}-10^{11.4}~M_\odot$ (HE quasar), respectively, 
which constitute the high-mass part of the local $M_\star$ function \citep[e.g.,][]{2013A&A...556A..55I}. 
These $M_\star$, for which Figure \ref{fig2} shows the distribution as a function of the redshift, 
are used to calculate $f_{\rm gas}$ directly. 

Furthermore, we also found one IR and three PG quasars 
in the optical $V$-band-based $M_\star$ catalog of \citet{2016ApJ...819L..27Z}. 
These will be appended in Section 3.2 as secondary data. 
The inclusion of this optical data does not affect our conclusion. 

\section{Relationship of $\lambda_{\rm Edd}$ and $f_{\rm gas}$}\label{sec3}
In this section, we investigate the relationship between $\lambda_{\rm Edd}$ and $f_{\rm gas}$ 
using either the indirect and direct method mentioned in Section 2. 

\subsection{Indirect method with $M_{\rm BH}$} 
\begin{figure*}
\begin{center}
\includegraphics[scale=0.5]{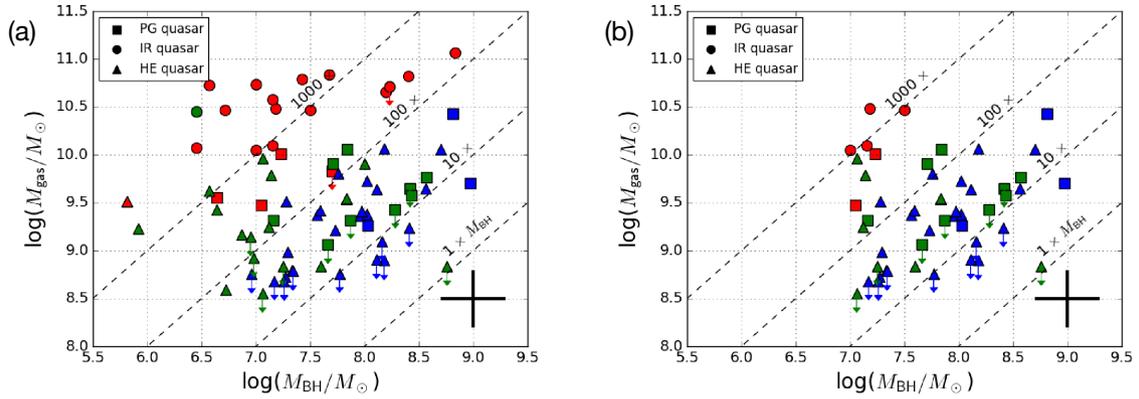}
\end{center}
\caption{
(a) Scatterplot of $M_{\rm BH}$ and $M_{\rm gas}$ for our sample of 
PG (squares), IR (circles), and HE (triangles) quasars. 
Colors indicate the Eddington ratio: 
$\lambda_{\rm Edd} > 1$ (red), $1 \geq \lambda_{\rm Edd} > 0.1$ (green), 
and $0.1 \geq \lambda_{\rm Edd}$ (blue). 
The dashed diagonal lines indicate that $M_{\rm gas}$ is 1, 10, 100, and 1000 times $M_{\rm BH}$. 
One can estimate $M_{\rm gas}/M_{\rm bulge}$ from these lines by assuming the local co-evolutionary relationship 
such as $M_{\rm BH} \simeq 0.005 \times M_{\rm bulge}$ \citep{2013ARA&A..51..511K}. 
The black cross shows the typical uncertainty (0.3 dex for both $M_{\rm BH}$ and $M_{\rm gas}$). 
(b) Same as (a), but for the subsample (see Section 3.1 for details). 
}
\label{fig3}
\end{figure*}

All of the quasars described in Section 2 (total number = 76) 
are plotted on the $M_{\rm BH}$--$M_{\rm gas}$ plane (Figure \ref{fig3}a). 
Here, we classify the samples into three groups based on their $\lambda_{\rm Edd}$: 
(i) the {\it super-$\lambda_{\rm Edd}$} group with $\lambda_{\rm Edd} > 1$, 
(ii) the {\it mid-$\lambda_{\rm Edd}$} group with $1 \geq \lambda_{\rm Edd} > 0.1$, 
and (iii) the {\it low-$\lambda_{\rm Edd}$} group with $0.1 \geq \lambda_{\rm Edd}$. 
While $\lambda_{\rm Edd} > 1$ defines super-Eddington objects, 
there is no physical motivation to set $\lambda_{\rm Edd} = 0.1$ as the threshold. 
Rather, these thresholds are adopted to clarify 
the dependence of $\lambda_{\rm Edd}$ on $f_{\rm gas}$. 
The figure also shows the fractions of $M_{\rm gas}$ with respect to $M_{\rm BH}$. 
Note that the actual $\lambda_{\rm Edd}$ of the super-$\lambda_{\rm Edd}$ group spans from 1.0 to 7.0. 

In Figure \ref{fig3}a, we observed a clear trend: 
most of the super-$\lambda_{\rm Edd}$ quasars were located in the region of $M_{\rm gas}/M_{\rm BH} \gtrsim 1000$, 
whereas the low-$\lambda_{\rm Edd}$ quasars were in $M_{\rm gas}/M_{\rm BH} \lesssim 10-100$. 
The mid-$\lambda_{\rm Edd}$ quasars tended to occupy the gap between the others. 
If we assume that the local co-evolutionary relationship 
\citep[e.g.,][]{2013ARA&A..51..511K} also holds for our quasar samples, 
i.e., $M_{\rm BH} \simeq 0.005 \times M_{\rm bulge}$ (here, $M_{\rm bulge}$ denotes galaxy {\it bulge} mass), 
the $M_{\rm gas}/M_{\rm BH}$ ratio can be transformed into $M_{\rm gas}/M_{\rm bulge}$. 
As such, we argue that the super-$\lambda_{\rm Edd}$ 
SMBHs occur in galaxies with a higher $M_{\rm gas}/M_{\rm bulge}$, 
i.e., in gas-rich systems, where we can also expect a high $f_{\rm gas}$. 

However, one would care about the bias on our sample selection 
that prefers optically bright objects as they were 
basically selected in flux-limited manners (Figure \ref{fig1}c): 
the SMBHs with lower $M_{\rm BH}$ could have been detected 
only when they have high enough $\lambda_{\rm Edd}$, 
while those with high $M_{\rm BH}$ could have been detected 
irrespective of their $\lambda_{\rm Edd}$. 
Thus, here we try another version of this method 
by only using those with $M_{\rm BH} \geq 10^7~M_\odot$ and $z \leq 0.15$ 
(Figure \ref{fig1}b: total number = 47). 
The low-mass cut is applied to ensure that objects with low-$\lambda_{\rm Edd}$ 
are not omitted in this {\it subsample}. 
Then the $M_{\rm BH}$--$M_{\rm gas}$ distribution is replot (Figure \ref{fig3}b). 
Here we again observe the same trend as in Figure \ref{fig3}a: 
high-$\lambda_{\rm Edd}$ SMBHs reside in gas-rich systems in a qualitative manner. 
Thus the trend we see in Figure \ref{fig3}a would not be affected by the selection bias significantly.

\subsection{Direct method with $M_\star$}
\begin{figure*}
\begin{center}
\includegraphics[scale=0.45]{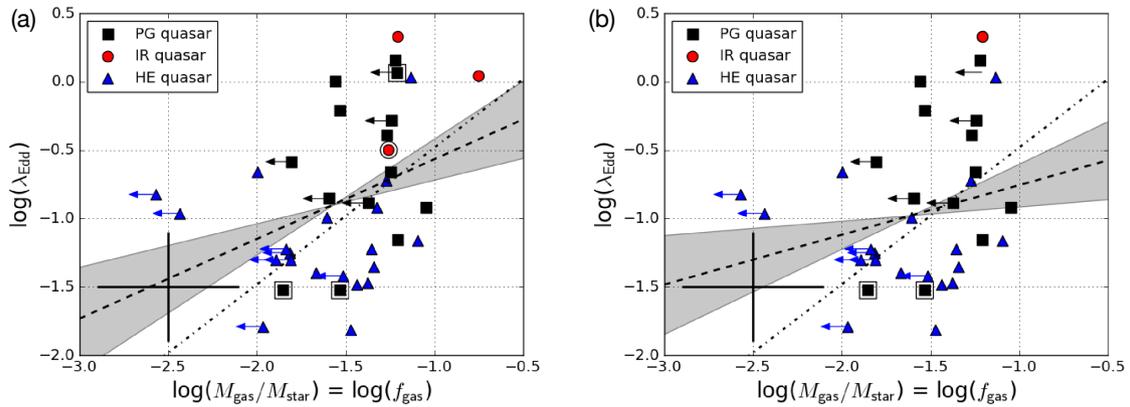}
\end{center}
\caption{
(a) $f_{\rm gas}$--$\lambda_{\rm Edd}$ correlation for the local quasar samples in the logarithmic space. 
In total 37 quasars (33 NIR-$M_\star$ samples and four optical-$M_\star$ samples) are plotted. 
The doubled symbols indicate the ancillary optical-$M_\star$ samples. 
The cross in the bottom-left indicates the expected typical uncertainty (0.4 dex for both $f_{\rm gas}$ and $\lambda_{\rm Edd}$). 
The dashed line shows the best-fit linear regression, log($\lambda_{\rm Edd}$) = (0.583 $\pm$ 0.265)$\times$log($f_{\rm gas}$) + (0.017 $\pm$ 0.416), 
with its uncertainty (gray-shaded region). 
See more details in Section 4. 
The dot-dashed line indicates the scaling of $\lambda_{\rm Edd}$ invoked in \citet{2012MNRAS.426..237H}, which has a slope of unity. 
(b) Same as (a), but for the subsample (see Section 3.2 for details). 
The best-fit linear regression here indicates 
log($\lambda_{\rm Edd}$) = (0.364 $\pm$ 0.259)$\times$log($f_{\rm gas}$) + ($-$0.393 $\pm$ 0.415). 
}
\label{fig4}
\end{figure*}

Next, we directly estimate $f_{\rm gas}$ by dividing $M_{\rm gas}$ by $M_\star$. 
In this manner, 33 quasars with the NIR-based $M_\star$ (hereafter NIR-$M_\star$ sample) 
and four ancillary quasars with $V$-band-based $M_\star$ (optical-$M_\star$ sample) 
described in Section 2 are plotted together in Figure \ref{fig4}a. 
The $f_{\rm gas}$ of the quasars (e.g., $\sim 10^{-1.5}$ in the PG quasar sample) 
agrees well with the model prediction at $z = 0$ by \citet{2014ApJ...794...69E}. 
Although a substantial fraction of the samples only has upper limits on $f_{\rm gas}$ 
due to the non-detection of $^{12}$CO(1--0) emissions, 
there could be a positive correlation between the two quantities. 

We thus performed the generalized Kendall's non-parametric correlation test 
to estimate the significance of the correlation, by using the IRAF
\footnote{IRAF is distributed by the National Optical Astronomy Observatory, 
which is operated by the Association of Universities for Research in Astronomy, Inc. 
under cooperative agreement with the National Science Foundation.} 
STSDAS\footnote{STSDAS is a product of the Space Telescope Science Institute, which is operated by AURA for NASA.} package. 
The program can treat mixed censored data, including censoring in the independent variable. 
The resultant Kendall's tau correlation coefficient for the NIR-$M_\star$ sample only is 0.49 and 
the probability that the data are uncorrelated is 3\%. 
The probability is reduced to 2\% if we use the full NIR- and optical-$M_\star$ samples, 
with almost the same coefficient of 0.49. 
With these values, we then suggest that $f_{\rm gas}$ 
and $\lambda_{\rm Edd}$ are (marginally) correlated. 

If we use the subsample data (i.e., $M_{\rm BH} \geq 10^7~M_\odot$ $\cap$ $z \leq 0.15$) only 
as was performed in Section 3.1, the resultant correlation becomes weaker 
(Figure \ref{fig4}b; correlation coefficient = 0.36 and the probability of no correlation = 10\%), 
although it still suggests a marginal and positive correlation. 
On the other hand, if we focus on the CO-detected objects (i.e., gas-rich systems) only, 
the data distribution is almost vertical both in Figure \ref{fig4}a and b. 
This would suggest the importance of expanding the sample to gas-poorer systems 
to depict the true galaxy distribution on this plane. 

\section{Discussion}\label{sec4}
Both the indirect and direct methods in \S 3 support the idea 
that $f_{\rm gas}$ and $\lambda_{\rm Edd}$ are positively correlated, 
although the latter method only gives a tentative support. 
From this, we argue that super-Eddington SMBHs, 
which are in the stage of rapid mass accumulation \citep{2004A&A...420L..23K}, 
tend to reside in relatively gas-rich galaxies. 
Our results are consistent with the observed decrease 
in $\lambda_{\rm Edd}$ over cosmic time \citep[e.g.,][]{2016ApJ...825....4T}, 
as $f_{\rm gas}$ of a galaxy is essentially a decreasing function 
with time, primarily due to star formation. 
This in turn implies that super-Eddington accretion could be a more common phenomenon at higher-$z$, 
where plenty of gas is expected. 
Indeed, such efficient accretion onto black hole seeds formed 
from the collapse of primordial stars \citep[e.g.,][for a recent review]{2017PASA...34...22G} 
are thought to be one of the most plausible paths for the formation of high(est)-$z$ 
SMBHs with $M_{\rm BH} \sim 10^9~M_\odot$ \citep[e.g.,][]{2018Natur.553..473B}. 

The $f_{\rm gas}$--$\lambda_{\rm Edd}$ correlation provides a clue 
to solving the cosmic downsizing problem too. 
For example, recent semi-analytic galaxy formation model of 
\citet{2012MNRAS.426..237H} reproduced the downsizing trend by modifying 
several of their physical recipes, including a {\it ceiling} of $\lambda_{\rm Edd}$ that depends on 
$f_{\rm gas}$, particularly at $z \leq 1$. 
\citet{2014ApJ...794...69E} also reproduced the trend 
with a decreasing amount of cold gas inside a galaxy with time, 
as well as with some other processes. 
In either model, a positive correlation of $\lambda_{\rm Edd}$ and $f_{\rm gas}$ is mandatory, 
which accords with our results (Figure \ref{fig4}) in a qualitative sense. 

However, the $f_{\rm gas}$--$\lambda_{\rm Edd}$ correlation observed here seems to be a bit shallower 
than the linear correlation in \citet{2012MNRAS.426..237H} with a logarithmic slope of unity: 
the two-dimensional Kaplan-Meier test using the same STSDAS package 
produces the linear regression in our data of 
log($\lambda_{\rm Edd}$) = (0.583 $\pm$ 0.265) $\times$ log($f_{\rm gas}$) + (0.017 $\pm$ 0.416), as shown in Figure \ref{fig4}a. 
Indeed, it is difficult to generate super-Eddington objects from this shallow best-fit relation. 
Further sensitive observations to reduce the samples with upper limits on $f_{\rm gas}$, 
increase those of high $f_{\rm gas}$ (e.g., $f_{\rm gas} \gtrsim 0.3$; these are particularly rare in the local universe), 
as well as a large number of samples to statistically overcome 
the time variable nature of the AGN continuum that hampers accurate 
determination of $\lambda_{\rm Edd}$ \citep[see][and references therein]{2011ApJ...737...26N}, 
are required to achieve a firmer $f_{\rm gas}$--$\lambda_{\rm Edd}$ relationship. 

The underlying factors driving the $f_{\rm gas}$--$\lambda_{\rm Edd}$ correlation 
could be (i) a co-evolutionary relationship 
and (ii) the positive correlation between $M_{\rm gas}$ 
and the accretion rate of AGN, $\dot{M}_{\rm BH}$ \citep[e.g.,][]{2011ApJS..195...23M,2017MNRAS.470.1570H}. 
Note that we invoked this idea in \S 3.1. 
Indeed, there is no clear correlation between $M_{\rm gas}$ itself 
(i.e., without dividing by $M_{\rm BH}$) and $\lambda_{\rm Edd}$ \citep{2017MNRAS.470.1570H}, suggesting the influence of (i). 
As expected in many mass accretion models \citep[e.g.,][]{2008ApJ...681...73K,2010MNRAS.407.1529H}, 
a key parameter controlling accretion is the gas mass surface density ($\Sigma_{\rm gas}$) 
at the central $\sim 100$ pc {\it circumnuclear disk (CND)} around an AGN. 
Indeed, \citet{2016ApJ...827...81I} observationally suggested the importance of the CND-scale $\Sigma_{\rm gas}$. 
The correlation of $M_{\rm gas}$ at a galaxy-scale and $\dot{M}_{\rm BH}$ implies 
that there is a type of scaling between $M_{\rm gas}$ and a CND-scale gas mass or $\Sigma_{\rm gas}$. 

In addition to the above, as a gaseous disk with higher $f_{\rm gas}$ is more susceptible to gravitational collapse 
\citep[e.g.,][]{2011ApJ...733..101G,2012MNRAS.424.1963S}, 
an induced series of gravitational instability will increase the mass inflow to the CND-scale or further inward, 
enhancing $\Sigma_{\rm gas}$ there, and eventually leading to a higher $\lambda_{\rm Edd}$. 
Violent gravitational instability is particularly important 
for triggering AGN in a high-$z$ universe \citep[e.g.,][]{2012ApJ...757...81B}, 
which even competes with another favored triggering mechanism, 
galaxy mergers/interactions \citep{2006ApJS..163....1H,2014A&A...569A..37M}. 
While it is difficult to investigate systematic differences in $\Sigma_{\rm gas}$ 
around super- and low-$\lambda_{\rm Edd}$ SMBHs with the current dataset, 
a systematic and statistical high-resolution study using 
the Atacama Large Millimeter/submillimeter Array (ALMA) should shed light on this.

\bigskip
\begin{ack}
We appreciate the anonymous referee for his/her careful reading 
and insightful comments and suggestions which improved this paper. 
We are grateful to A. Schulze for providing his HE quasar data. 
T.I. greatly appreciates M. Imanishi, T. Kawamuro, and A. Schulze at NAOJ for fruitful discussion. 
T.I. is supported by JSPS KAKENHI Grant Number 17K14247. 
\end{ack}

\end{document}